\documentclass[a4paper,11pt]{article}

\usepackage{color}
\usepackage{adjustbox}
\usepackage{graphicx}
\usepackage{float}
\usepackage{multirow}
\usepackage[mathscr]{eucal}
\usepackage{subfigure}
\usepackage{epstopdf}
\usepackage{caption}
\usepackage{verbatim}
\usepackage{amsmath}
\usepackage{natbib}
\usepackage{lineno}

\usepackage{amsthm}
\usepackage{amsfonts}
\usepackage{a4}
\usepackage[utf8]{inputenc} 
\usepackage{times}
\usepackage{here}
\usepackage[usenames,dvipsnames,svgnames,table]{xcolor}
 \usepackage{hyperref}

\definecolor{darkcandyapplered}{rgb}{0.64, 0.0, 0.0}
\definecolor{darkcerulean}{rgb}{0.03, 0.27, 0.49}
\makeatletter
\newcommand{\manuallabel}[2]{\def\@currentlabel{#2}\label{#1}}

\newcommand{\X}{\mathbf{{X}}}

\newcommand{\Z}{\mathbf{{Z}}}

\newcommand{\bSigma}{{\mbox{\boldmath $\Sigma$}}}
\newcommand{\bbeta}{{\mbox{$\boldsymbol{\beta}$}}}

\newcommand{\btheta}{{\mbox{\boldmath $\theta$}}}
\newcommand{\bphi}{{\mbox{\boldmath $\phi$}}}
\newcommand{\bb}{{\mbox{$\boldsymbol{b}$}}}

\newcommand*{\Scale}[2][4]{\scalebox{#1}{$#2$}}%

\makeatother

\begin{document}
\title{Bayes factors  for longitudinal model assessment via power posteriors}

\author{Gabriel Calvo$^{1}$, Carmen Armero$^{1}$, Luigi Spezia$^{2}$, and Maria Grazia Pennino$^{3}$}
\maketitle
\noindent $^1$ Department of Statistics and Operations Research,
             Faculty of Mathematics, Univer-\\ 
             \hspace*{1cm}sitat de Val\`encia, Spain.
     {\textbf{gabriel.calvo@uv.es}},
             {carmen.armero@uv.es} \\
$^2$ Biomathematics \& Statistics Scotland, Aberdeen, UK.
{luigi.spezia@bioss.ac.uk} \\  
$^3$ Instituto Espa\~nol de Oceanograf\'ia (IEO, CSIC), Centro Oceanogr\'afico de Vigo,\\ 
             \hspace*{1cm} Spain. {grazia.pennino@ieo.es} 

\begin{abstract}
 Bayes factor,  defined as the ratio of the marginal likelihood functions of two competing models, is the natural Bayesian procedure for model selection.  Marginal likelihoods  are usually computationally demanding and complex. This scenario is particularly cumbersome    in linear mixed  models (LMMs) because   marginal likelihood functions involve integrals of large dimensions   determined by the number of parameters and the number of random effects, which in turn increase with the number of individuals in the sample. The power posterior  is an attractive proposal in the context of the Markov chain Monte Carlo  algorithms that allows  expressing   marginal  likelihoods  as    one-dimensional  integrals over the unit range. This paper explores the use of power posteriors in  LMMs and  discusses their behaviour through two simulation studies  and a real data set on European sardine landings in the Mediterranean Sea.
\end{abstract}

\section{Introduction}

 Model selection is a key issue in   parametric statistics that has generated a large scientific literature. Several proposals have been made from different perspectives to carry out  model comparison. This is a  multifaceted topic with a high philosophical content that will undoubtedly continue to improve scientific knowledge in the future.  

 Bayesian methodology uses a conception of probability that allows assigning probability distributions to any kind of quantity with uncertainty, in particular parameters, hyperparameters, and probabilistic models. The posterior distributions of these quantities are the basis of the natural Bayesian procedure for model selection through the  posterior distribution for each of the candidate models. When comparing two competing models, the ratio of their corresponding posterior distributions (posterior odds) is obtained through the product of the prior odds and the Bayes factor, the latter defined in terms of the ratio of the marginal likelihood  of each of the two models \citep{Kass1995, Berger1996}.  This is a conceptually simple and powerful procedure with a key limitation in that it is indeterminate when working with improper prior distributions, i. e., those distributions that do not integrate to unity. 
 
Longitudinal data are observations of one or more variables measured over time
 on each of the individuals in the study. They include observations between and within individuals that allow the assessment of general patterns of the target population   as well as specific individual characteristics. They are multivariate, clustered, and repeated measures data.  Data among individuals are commonly assumed to be independent, whilst the repeated measurements within each subject are correlated \citep{hedeker2006longitudinal}. Linear mixed effects models (LMMs) \citep{Laird1982, pinheiro2006mixed} constitute a flexible and powerful tool for the analysis of longitudinal data within  the normal distribution framework.
 
 The computation of  marginal likelihoods for assessing Bayes factors can be demanding and complex. In fact, this situation is  exacerbated in models  such as LMMs. The marginal likelihood function for these models involves the integral with respect to the prior  distribution of the conditional likelihood function that depends on both the parameters and the random effects. The dimension of this integral    increases  with  the  size  of  the random effects set that, in turn, increases with the number of individuals in the sample. 
 
 The challenge in calculating the marginal likelihoods in these  models has produced a large literature and a fruitful scientific debate, which in turn has generated different proposals for its computation. All of them have interesting properties, but none of them has managed to close the issue definitively, mainly due to  computational problems. Although it is not our aim to present in this paper an exhaustive list of all of the proposals, we would like to comment very briefly on some of the most popular ones. One of the first procedures to estimate marginal likelihoods was through the Laplace's method \citep{Tierney1986}.  Subsequently, \cite{newton1994approximate} expressed marginal likelihoods through the posterior harmonic mean of the likelihood, a simple but computationally unstable approach \citep{Raftery2007}.   \cite{chib1995marginal} and \cite{chib2001marginal} developed an algorithm which computes marginal likelihoods from the Markov chain Monte Carlo (MCMC) outputs by making use of additional iterations. Importance sampling ideas have been proposed for estimating the evidence such as  annealed importance sampling  \citep{Neal2001} or bridge sampling  \citep{MengWong1996}. Nested sampling proposed by \cite{Skilling2006} represents  marginal likelihoods in terms of  one-dimensional integral over $[0,1]$. This is an interesting approach based on simulated values from the prior distribution   subject to constraints in the conditional likelihood but it involves  many challenges when working with multidimensional models and prior distributions poorly informative. 
 
 The power posterior (\cite{Lartillot2006}, \cite{Friel2008}, and \cite{Friel2012}) is a proposal   developed in the context of MCMC algorithms in which the logarithm of the marginal likelihood is evaluated numerically through the path sampling algorithm of \cite{Gelman1998}, also called thermodynamic integration, to compute ratios of normalising constants.  

Our paper generalises the use of power posterior for approximating marginal likelihoods into complex models which include random effects and serial correlation terms in their conditional formulations. In particular, we  extend the use of power posteriors to LMMs, although it can be easily  generalised to any other models with random effects,  and we discuss its behaviour  through two simulated studies  and a real study on European sardine (\textit{Sardina pilchardus}, Walbaum, 1792) landings in the Mediterranean Sea. The data is available at \url{https://github.com/gcalvobayarri/Bayes_factor_longitudinal_models.git}.

This paper is organised as follows.   Section 2 introduces the basic Bayesian linear mixed models (BLMMs) and describes two popular generalisations based on autoregressive terms.  Section 3 reviews   Bayes factors and marginal likelihoods and presents the power posterior for models with random effects.  Section 4  assesses the behaviour of the power posterior in two simulated longitudinal studies.    The first one deals with data generated from a simple LMM model that competes with three basic models. The second databank was simulated from a model with a high level of complexity, in the sense that it includes three types of random variation: random effects, an autoregressive term and normally distributed measurement errors. This model is compared with two simpler competing models with only two of the three sources of variation discussed above.   Section 5 is devoted to the selection of a Bayesian longitudinal model to analyse sardine fisheries in different countries of the Mediterranean Sea. The paper  concludes with a small discussion emphasising the usefulness of the power posterior methodology in the calculation of marginal likelihoods for comparing LMMs,  and  highlights the most interesting results of the three studies carried out in the work.  Finally, Appendix collects the full conditionals of the Gibbs sampling  for all of the models in the paper.

\section{Bayesian  longitudinal linear mixed   models}\label{sec:joint_models}

Let $\boldsymbol{y}_i = (y_{i1}, \ldots, y_{in_i})^{\prime}$ be the random vector describing the response of individual $i$, $i=1,\dots,N$, recorded at  times $\boldsymbol{t}_i = (t_{i1}, \ldots, t_{in_i})^{\prime}$, and consider $\boldsymbol{y}  = (\boldsymbol y_{i}, \ldots, \boldsymbol y_{N})^{\prime}$. A BLMM for    $\boldsymbol{y}$ is specified via  the joint probability distribution   
\begin{align}  
     f(\boldsymbol{y},\btheta,\bphi) = & \,f(\boldsymbol{y} \mid \btheta,\bphi)\,f(\bphi|\btheta)\,\pi(\btheta) \nonumber\\
     = & \,\big(\prod_{i=1}^{N}\,f(\boldsymbol{y_i} \mid \btheta,\bphi_i)\big)\,\big(\prod_{i=1}^{N}\,f(\bphi_i \mid \btheta)\big)\,\pi(\btheta),   \label{eqn:joint} 
\end{align}

\noindent where  $f(\boldsymbol{y}_i \mid \btheta,\bphi_i)$  is the conditional distribution of $\boldsymbol y_i$ given the vector $\bphi_i$ 
of random effects associated with individual $i$, with  $\bphi=(\bphi_1,\ldots, \bphi_N)^\prime$, and  the vector $\btheta$   of parameters and hyperparameters of the model; $f(\bphi_i \mid \btheta)$  is the conditional distribution    of $\bphi_i$ given $\btheta$; and  $\pi(\btheta)$  the prior distribution  for $\btheta$.

The simplest BLMM assumes the following conditional normal      distribution for   $\boldsymbol{y}_i$:
\begin{equation}
\label{eqn:normal1}
     f(\boldsymbol{y}_i \mid \btheta,\bphi) = \mathcal{N}(\boldsymbol \mu_i = \X_i \bbeta + \Z_i \bb_i, \, \bSigma_i=\sigma^2 \,I_{n_{i}}),
\end{equation}
 
 \noindent where $\X_i$ and $\Z_i$ are the design matrices for the fixed effects $\bbeta$ and the  random effects $\boldsymbol b_i$, respectively, and $\bSigma_i$ is the variance-covariance matrix defined in terms of the  identity matrix $I_{n_{i}}$ of size $n_i$ and the common variance $\sigma^2$. The vector of random effects $\boldsymbol \phi$ in (\ref{eqn:joint}) is here     $\boldsymbol b=(\boldsymbol b_1, \ldots, \boldsymbol b_N)^{\prime}$ whose components are assumed conditional i.i.d. 
 $(\bb_i \mid \bSigma_b) \sim \mathcal{N}(0, \bSigma_b)$. This conditional model (\ref{eqn:normal1}) implies conditional independence  not only between the observations of different individuals but also of all observations of the same individual. 
 
Common generalisations of (\ref{eqn:normal1}) when the number of observations per individual is large include temporal elements that account for serial correlation \citep{diggle2002analysis}, frequently in terms of autoregressive processes in the conditional mean $\boldsymbol \mu_i$ or in the conditional variance-covariance matrix $\bSigma_i$. In this sense, we consider two different modelling approaches based on a first order autoregressive ($AR(1)$) process.  The first proposal   introduces an  $AR(1)$ term in the random measurement \citep{chi1989models, hedeker2006longitudinal} which, as a result, reformulates the $(j,l)$ element of $\bSigma_i$ in (\ref{eqn:normal1}) as follows:

\begin{equation}
\label{eqn:covariance} \mbox{Cov}(y_{ij}, y_{il} \mid \boldsymbol \theta, \boldsymbol \phi_i)=  \frac{\sigma^2}{1-\rho^2} \, \rho^{|t_{ij}-t_{il}|},
\end{equation}
 
\noindent where $\rho$ is the coefficient of the autoregressive term of a stationary $AR(1)$. Because of the stationarity condition, $\rho$ is also equal to the conditional  autocorrelation at lag 1. More generally, $\rho^{h}$ is the conditional autocorrelation at lag $h$ or $-h$, with $h\geq0$. This model maintains the conditional independence between individuals but not between the observations of the same individual. 
Furthermore, a reformulation of this model is possible by conditioning the response of individual $i$ in the $j$-th measurement, $y_{ij}$ on their $(j-1)$-th measurement as follows
\begin{equation}
\label{eqn:normal_ar1}
     f({y}_{ij} \mid {y}_{ij-1}, \btheta,\bphi) = \mathcal{N}( \mu_{ij} + \rho({y}_{ij-1}-\mu_{ij-1}), \, \sigma^2), \,\,j=2,\ldots, t_{in_{i}}
\end{equation}

\noindent being $f({y}_{i1} \mid \btheta,\bphi) = \mathcal{N}( \mu_{i1} , \, \sigma^2/(1-\rho^2))$.

The second proposal  introduces a latent autoregressive element  
$\boldsymbol w_i=(w_i(t_{i1}), \ldots, w_i(t_{in_{i}}))^{\prime}$   in the conditional mean $\boldsymbol \mu_i$ of (\ref{eqn:normal1}) as follows:

\begin{equation}
\label{eqn:normal2}
       \boldsymbol \mu_i = \X_i \bbeta + \Z_i \bb_i+\boldsymbol w_i, 
\end{equation}

\noindent where each $w_i(t_{ij})$, $j=2,\ldots, n_i$, is a realisation at time $t_{ij}$  from a Gaussian process with mean $\rho \, w_i(t_{i,j-1})$ and variance $\sigma^2_w$, where $(w_i(t_{i1}) \mid \rho, \sigma_w)  \sim \mathcal{N}(0, \sigma^2_w / (1-\rho^2))$  \citep{diggle2002analysis}, that is $\boldsymbol w_i$ is a vector of time correlated noises.  This model augments the vector of random effects associated with individual $i$ to  $\boldsymbol \phi_i=(\boldsymbol b_i, \boldsymbol w_i)$. The two elements in $\boldsymbol \phi_i$ are conditionally independent given $ \boldsymbol \theta$.


 The full specification of the Bayesian model is completed with the  elicitation of  a prior distribution $\pi(\boldsymbol \theta)$ for
 $\boldsymbol \theta $. This includes parameters in $\boldsymbol \beta$ and   $\bSigma_i$ as well  as hyperparameters in the covariance-matrix  $\bSigma_b$ and in the autoregressive terms.
 
Let $\boldsymbol y_{obs}$ denote the vector of data. On the basis of the information provided by $\boldsymbol y_{obs}$, we can calculate the posterior distribution of  $(\boldsymbol \theta, \boldsymbol \phi)$. This  is the most important element of the Bayesian inferential process from which we can derive   posterior distributions for relevant outcomes of the problem. It is obtained through the Bayes' theorem according to 
 
 \begin{equation}
\label{eqn:posterior}
\pi(\boldsymbol \theta, \boldsymbol \phi  \mid \boldsymbol y_{obs} ) \propto 
 f(\boldsymbol y_{obs} \mid \boldsymbol \theta, \bphi) \,f(\boldsymbol \phi \mid \boldsymbol \theta) \, \pi(\boldsymbol \theta),
 \end{equation}
 
\noindent where    $f(\boldsymbol y_{obs} \mid \boldsymbol \theta, \bphi)$ represents now    the likelihood function of  $(\boldsymbol \theta, \boldsymbol \phi)$ given $\boldsymbol y_{obs}$ which  is expressed as the product of each individual's contribution to the likelihood:
\begin{equation}
\label{eqn:likelihood}
     f(\boldsymbol y_{obs}  \mid \boldsymbol \theta, \bphi) = \prod_{i=1}^{N}\, f(\boldsymbol{y}_{obs_{i}} \mid \btheta,\bphi_i) = \prod_{i=1}^{N}\, \frac{\text{exp}\big(- \frac{1}{2}(\boldsymbol{y}_{obs_{i}} - \boldsymbol \mu_i )^{\prime} \, \bSigma_i \, (\boldsymbol{y}_{obs_{i}} - \boldsymbol \mu_i )     \big)}{\sqrt{(2\pi)^{n_i}|\bSigma_i |} }.
\end{equation}
 
  Note that   the  posterior  distribution  $\pi(\boldsymbol \theta, \boldsymbol \phi  \mid \boldsymbol y_{obs})$ in (\ref{eqn:posterior}) involves $\boldsymbol \theta$ and $\boldsymbol \phi$ together. We have chosen to express the   prior  information on them  as $f(\boldsymbol \phi \mid \boldsymbol \theta) \, \pi(\boldsymbol \theta)$ and not as $\pi(\boldsymbol \theta,  \boldsymbol \phi)$. This  is a philosophically debatable topic  and it is not our intention to enter into this issue here.  We have simply chosen that expression because we recognise it as the most standard one.   
 

\section{Bayes factors and power posteriors}

The marginal or predictive density $m_k(\boldsymbol y)$ for $\boldsymbol y$ when considering a BLMM model   $\mathcal M_k$ with parameter and hyperparameter  $\btheta_k$ and random effects  $\boldsymbol \phi_k$ is defined as
\begin{align} 
\label{eq:marginal_likelihood}
  m_k(\boldsymbol y) & =\int \, f( \boldsymbol y, \boldsymbol \btheta_k, \boldsymbol \phi_k) \, \mbox{d}(\btheta_k,\, \phi_k) \nonumber \\
  & =\int \, f( \boldsymbol y \mid \boldsymbol \btheta_k, \boldsymbol \phi_k) \, f(\boldsymbol \phi_k \mid \boldsymbol \btheta_k  )  \, \pi(\boldsymbol \btheta_k ) \, \mbox{d}(\btheta_k,\, \phi_k). 
\end{align}

This  distribution evaluated on the data, $m_k(\boldsymbol y_{obs})$, is known  by different terms in the literature such as marginal likelihood \citep{newton1994approximate}, predictive distributon \citep{gelfand1994bayesian}, marginal probability \citep{Kass1995}, predictive probability \citep{Lartillot2006} or evidence \citep{Friel2012}. It can be interpreted as the support provided by the data in favour of  model  $\mathcal M_k$. 

The main tool used for choosing between two models, $\mathcal M_1$ and $\mathcal M_2$, in the Bayesian methodology is the Bayes factor \citep{Kass1995, Berger1996} of model $\mathcal M_1$ against model $\mathcal M_2$. It is defined    as follows

$$B_{12}=\frac{m_1(\boldsymbol y_{obs})}{m_2(\boldsymbol y_{obs})},$$

\noindent and measures the strength with which the data support  $\mathcal M_1$ with regard to  $\mathcal M_2$. 

In the following, we will  generalise the power posterior    to the case of longitudinal models involving not only parameters and hyperparameters  $\boldsymbol \theta$  but also random effects  $\boldsymbol \phi$. Accordingly, we define the power posterior of $(\boldsymbol \theta, \boldsymbol \phi)$ as 
\begin{equation}
   \pi_\tau(\btheta, \boldsymbol \phi | \boldsymbol y_{obs} ) \propto  f( \boldsymbol y_{obs} \mid \boldsymbol \btheta, \boldsymbol \phi)^\tau \, f(\boldsymbol \phi \mid \boldsymbol \theta) \,\pi(\btheta),  
\end{equation}
 \noindent where $\tau \in [0,1]$ is an auxiliary temperature variable that modulates the effect of the likelihood. Let the ``power marginal likelihood'' be
 
     \begin{equation}\label{eq:marg_pow}
  m(\boldsymbol y_{obs} \mid \tau)=  \int      f( \boldsymbol y_{obs} \mid \boldsymbol \btheta, \boldsymbol \phi)^\tau \, f(\boldsymbol \phi \mid \boldsymbol \theta) \,\pi(\btheta) \, \mbox{d}(\boldsymbol \theta, \boldsymbol \phi);  
     \end{equation}

\noindent when $\tau=0$, $m(\boldsymbol y_{obs} \mid \tau=0)= \mbox{$\int$}  f(\boldsymbol \phi \mid \boldsymbol \theta) \,\pi(\btheta) \, \mbox{d}(\boldsymbol \theta, \boldsymbol \phi) =1$, whereas when $\tau=1$,  $m(\boldsymbol y_{obs} \mid \tau=1)$ equals the marginal likelihood $m(\boldsymbol y_{obs})$, our target quantity. Moreover, derived from the expression (\ref{eq:marg_pow}) we obtain the following result: 
\begin{align*}
  \mbox{E}_{(\boldsymbol \theta, \boldsymbol \phi \mid  \boldsymbol y_{obs}, \tau)}(&\mbox{log}\,(f( \boldsymbol y_{obs} \mid \boldsymbol \btheta, \boldsymbol \phi)))=  \int\, \mbox{log}\,(f( \boldsymbol y_{obs} \mid \boldsymbol \btheta, \boldsymbol \phi)) \, \pi_{\tau}(\boldsymbol \theta, \boldsymbol \phi \mid \boldsymbol y_{obs}) \, \mbox{d}(\boldsymbol \theta, \boldsymbol \phi )  \\
   = & \, \frac{1}{m(\boldsymbol y_{obs} \mid \tau)} \, \int\, \mbox{log}\,(f( \boldsymbol y_{obs} \mid \boldsymbol \btheta, \boldsymbol \phi)) \, \,f( \boldsymbol y_{obs} \mid \boldsymbol \btheta, \boldsymbol \phi)^{\tau} \, f(\boldsymbol \phi \mid \boldsymbol \theta) \, \pi(\boldsymbol \theta) \,\mbox{d}(\boldsymbol \theta, \boldsymbol \phi )  \\
   = & \, \frac{1}{m(\boldsymbol y_{obs} \mid \tau)} \,\int \Big[\,   \,\frac{\mbox{d}}{\mbox{d} \tau} \, f(\boldsymbol y_{obs} \mid \boldsymbol \theta, \boldsymbol \phi)^{\tau} \, f(\boldsymbol \phi \mid \boldsymbol \theta) \, \pi(\boldsymbol \theta)\Big] \,\mbox{d}(\boldsymbol \theta, \boldsymbol \phi ) \\
    = & \,\frac{1}{m(\boldsymbol y_{obs} \mid \tau)} \, \frac{\mbox{d}}{\mbox{d} \tau} \,\Big[\int\,   \,f(\boldsymbol y_{obs} \mid \boldsymbol \theta, \boldsymbol \phi)^{\tau} \, f(\boldsymbol \phi \mid \boldsymbol \theta) \, \pi(\boldsymbol \theta) \,\mbox{d}(\boldsymbol \theta, \boldsymbol \phi ) \Big]\\
     = & \,\frac{1}{m(\boldsymbol y_{obs} \mid \tau)} \, \frac{\mbox{d}}{\mbox{d} \tau} \, m(\boldsymbol y_{obs} \mid \tau) \\
     = & \, \frac{\mbox{d}}{\mbox{d} \tau} \,\mbox{log}\,(m(\boldsymbol y_{obs} \mid \tau)). 
    \end{align*}

\noindent As a consequence, we have
\begin{equation}
 \int_{0}^{1}\, \mbox{E}_{(\boldsymbol \theta, \boldsymbol \phi \mid  \boldsymbol y_{obs}, \tau)}(\mbox{log}\,(f(\boldsymbol y_{obs} \mid \boldsymbol \theta, \boldsymbol \phi))\, \mbox{d}\tau  = \mbox{log} \, \Big( \frac{m(\boldsymbol y_{obs} \mid \tau=1)} {m(\boldsymbol y_{obs} \mid \tau=0)} \Big ) = \mbox{log}\,(m(\boldsymbol y_{obs})).
    \end{equation}
    
 This expression in the log-scale by means of the expectations provides numerical stability to the estimation of the marginal likelihood \citep{Friel2008}. 
 
For any value $\tau \in \left[ 0, \,1\right] $,  an estimate of the expectation
${E}_{(\boldsymbol \theta, \boldsymbol \phi \mid  \boldsymbol y_{obs}, \tau)}(\mbox{log}
\,(f(\boldsymbol y_{obs} \mid \boldsymbol \theta, \boldsymbol \phi)))$ and an estimate of the variance 
${V}_{(\boldsymbol \theta, \boldsymbol \phi \mid  \boldsymbol y_{obs}, \tau)}(\mbox{log}
\,(f(\boldsymbol y_{obs} \mid \boldsymbol \theta, \boldsymbol \phi)))$ are obtained by
using the MCMC draws generated from the power posterior $\pi_\tau(\btheta, \boldsymbol \phi | \boldsymbol y_{obs} )$. Running $M+1$ MCMC algorithms with a
temperature ladder such as $0=\tau _{0}<\tau _{1}<\ldots   <\tau
_{M}=1$, \cite{Friel2008} provides an estimate  of $\mbox{log}\,m(\boldsymbol y_{obs})$ by
applying the trapezoidal rule, which in our case will remain as   follows:
\begin{align} \label{eqn:trapeci}
\mbox{log}&(m(\boldsymbol y_{obs}))   \approx \sum_{m=0}^{M-1} \,  (\tau
_{m+1} -\tau _{m}) \times \nonumber \\
    &\times \frac{1}{2}\Big( E_{(\boldsymbol{\theta}, \boldsymbol{ \phi \mid}  \boldsymbol y_{obs}, \tau_{m})}( \mbox{log} \,(f(\boldsymbol y_{obs} \mid \boldsymbol \theta, \boldsymbol \phi)) +
E_{(\boldsymbol{\theta}, \boldsymbol{ \phi \mid}  \boldsymbol y_{obs}, \tau_{m+1})}( \mbox{log} \,(f(\boldsymbol y_{obs} \mid \boldsymbol \theta, \boldsymbol\phi) ) \Big)
\end{align}

The temperature ladder from $\tau =0$ to $\tau =1$ defines a path from the prior distribution to the posterior distribution, hence the
name of path sampling (\cite{Gelman1998}).

\section{Simulation studies}

We explore below the behaviour of the power posterior for approximating the evidence in Bayesian longitudinal models.  Two different simulation studies are conducted to evaluate the capability of the power posterior to identify the   model which  generated the data within a set of potential competitors. In the first study, we work with a small data set generated from a simple model with  random effects and normally distributed measurement errors. In the second case,  we have a larger dataset generated from a model with more complexity we have a larger dataset generated by a model with more complexity including random effects, measurement errors and an autoregressive term.

\subsection{Study 1: a balanced longitudinal data set}

We assume a scenario based on  a small  balanced set of simulated longitudinal data generated from
a simple model,  approximate the evidence of this model for the simulated data,   and compare it with the evidence of three other competing longitudinal models for the same data. 

\begin{figure}[H]
	\centering
	  \includegraphics[width=120mm]{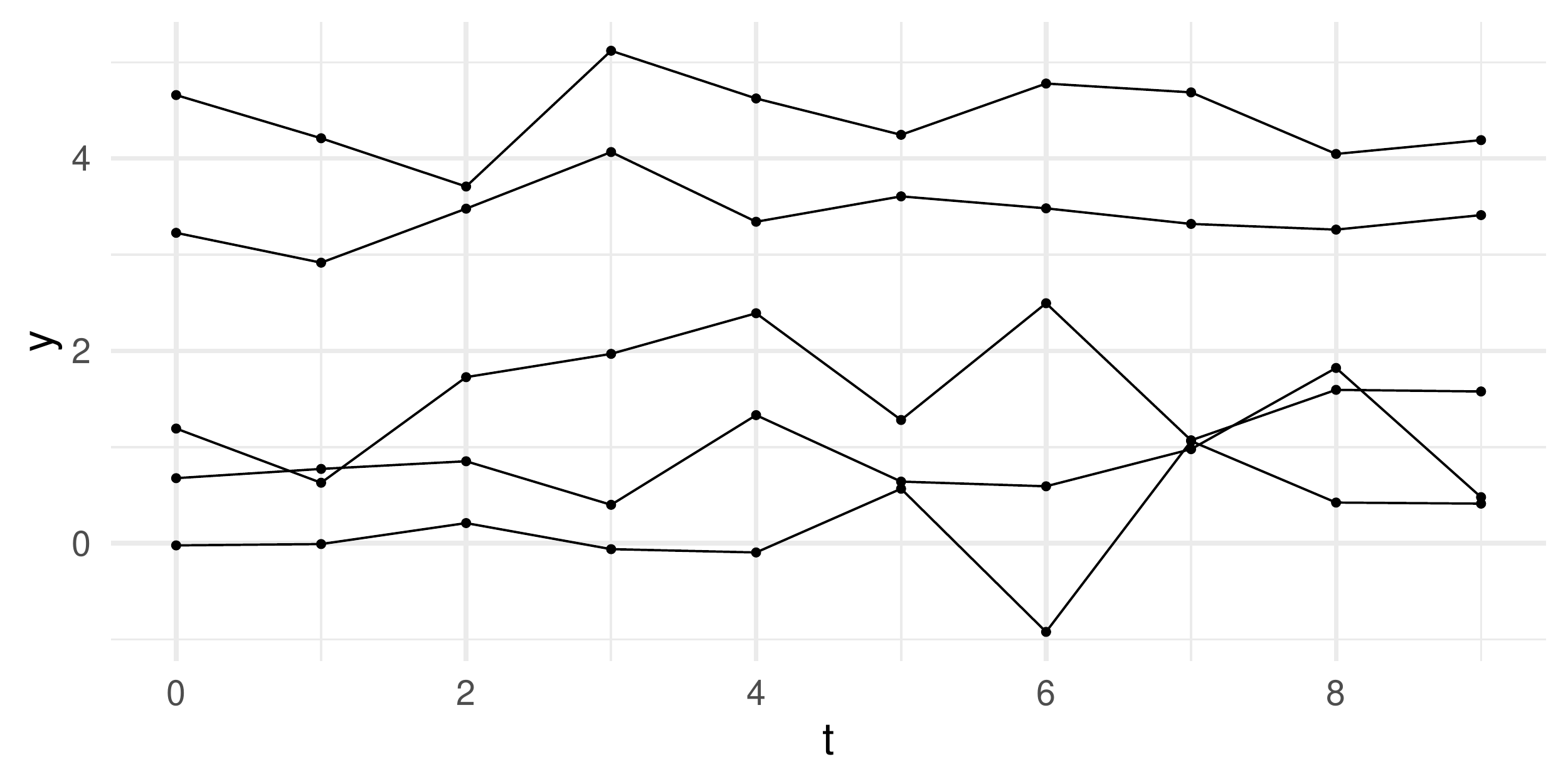} 
	 	\caption{Spaghetti plot of the simulated longitudinal data from the first study, 5 individuals are considered in a complete balanced design for times $\{0, 1, \ldots, 9\}$.} \label{fig:spaghetti_plots_sim}
\end{figure}

We consider the LMM (\ref{eqn:normal1}) with   design matrices   $\X_i=\Z_i=\boldsymbol{1} $, where $\boldsymbol 1$ is a $n_i$-vector whose components are all the unity 1. We represent by   $\boldsymbol 1 \, \beta_0 $  the   common intercept and by  $\boldsymbol 1 \, b_{0i}$ the individual random intercepts, which are conditionally normally distributed, $(b_{0i} \mid \sigma_0^2) \sim \mathcal{N}(0, \sigma_0^2)$. We set the values  $\beta_0 = 2, \,\,\sigma= 0.5$, and  $\sigma_0 = 1.5$ for the parameters and hyperparameter of the model,  and generate data for $n_i=10$ response values corresponding to $N=5$ individuals in a complete balanced design for times $\{0, 1, \ldots, 9\}$ (see Figure \ref{fig:spaghetti_plots_sim}). 

Four competing longitudinal models $\mathcal M_1, \mathcal M_2, \mathcal M_3 \text{ and } \mathcal M_4$, ordered from least to most complex,    are taken into account for analysing the simulated data, i.e.,

\begin{align} 
 \mathcal M_1: &  \,\,f( y_{it} |\btheta_1) = \mathcal{N}(\beta_0,\sigma^2),  \nonumber\\
 \mathcal M_2: &  \,\,f( y_{it} |\btheta_2, \bphi_2) = \mathcal{N}(\beta_0 + b_{0i},\sigma^2),  \nonumber \\
 \mathcal M_3: &  \,\,f( y_{it} |\btheta_3, \bphi_3) = \mathcal{N}(\beta_0 + b_{1i}t,\sigma^2),  \nonumber \\
 \mathcal M_4: &  \,\,f( y_{it} |\btheta_4, \bphi_4) = \mathcal{N}(\beta_0 + b_{0i} + b_{1i}t,\sigma^2). \label{eqn_model_simu1}
\end{align} 

\noindent The model that generated the  data is $\mathcal M_2$. It includes a random effect associated with the intercept of each individual ($b_{0i}$)  and a normal measurement error. Model  $\mathcal M_1$, with only normal measurement errors, is the simplest model.  $\mathcal M_3$  and $\mathcal M_4$ includes normal measurement errors, $\mathcal M_3$ also  includes random slopes associated to individuals ($b_{1i}$),  and $\mathcal M_4$ considers  random   intercepts and random slopes. 

All four models assume conditional independence between individuals and within observations  from the same individual   as well as homogeneity of variances. Individual random effects $b_{0i}$ and random slopes $b_{1i}$  are mutually independent and conditionally  normally distributed  as $(b_{0i} \mid \sigma_0^2) \sim \mathcal{N}(0, \sigma_0^2)$ and $(b_{1i} \mid \sigma_1^2) \sim \mathcal{N}(0, \sigma_1^2)$. 

The elicitation of the subsequent prior distribution $\pi(\btheta)$ in all models is based on both prior independence among the parameters and a noninformative prior scenario: a normal distribution for the common intercept $\pi(\beta_0)=\mathcal{N}(0, 10^2)$ and wide uniforms for the standard deviation parameters $\pi(\sigma)=\pi(\sigma_0)=\pi(\sigma_1)=\mbox{U}(0, 10)$. 


For the computation of the evidence  via the power posterior in each of the four models in (\ref{eqn_model_simu1}), the temperature variable $\tau$ was discretised  according to  $0=\tau_0 < \tau_1 < \dots < \tau_{199}=1$ with $\tau_r = (r / 199)^5$ and $r=0,\ldots,199$. This temperature ladder ensures that a high proportion of values are close to $0$, improving the convergence of the algorithm. Each power posterior is computed by the Gibbs sampling, and implemented in the R environment, version 4.0.5, \citep{Rprogram}  running a chain with $50,000$ iterations for each discrete value of $\tau$. Then, we calculate an estimate of the model evidence applying the trapezoidal rule (\ref{eqn:trapeci}). In addition, to quantify the variability of the process, we repeat the algorithm ten times per model.

\begin{table}[h]
 \centering
\caption{Mean (standard deviation) of the ten replicates of the approximate log evidence for each model in (\ref{eqn_model_simu1}) computed by means of the power posterior method.}\label{tab:res_simulations}
\begin{tabular}{cccc}
\noalign{\hrule height 1pt}
              $\mathcal M_1$    & $\mathcal M_2$      & $\mathcal M_3$      & $\mathcal M_4$                              \\ \cline{1-4} \multicolumn{1}{r}{$-103.01(0.14)$} & \multicolumn{1}{r}{$-51.63\,(0.11)$} & \multicolumn{1}{r}{$-91.83\,(0.14)$} & \multicolumn{1}{r}{$-68.21\,(0.32)$} \\
\noalign{\hrule height 1pt}
\end{tabular}
\end{table}

Table \ref{tab:res_simulations} shows for each model  the mean and the standard deviation of the ten values of the approximate logarithm of the evidence. The evidence for each of the four models is clearly ordered. The model with the highest log evidence ($-51.63$) is actually the true sampling model $\mathcal M_2$, followed by the complete mixed linear model  $\mathcal M_4$  which includes a common population intercept and two types of individual random effects, $b_{0i}$ and $b_{1i}$. In contrast, the model with the lowest log marginal likelihood value ($-103.01$) is the fixed effects model $\mathcal M_1$. The standard deviation for  $\mathcal M_4$ is the highest ($0.32$), possibly because it is the most complex  of the four models. Note also that the variability associated with the replicate process is  relatively low in all cases.

\subsection{Study 2: an unbalanced longitudinal data set with serial correlation}

\begin{figure}[H]
	\centering
	  \includegraphics[width=120mm]{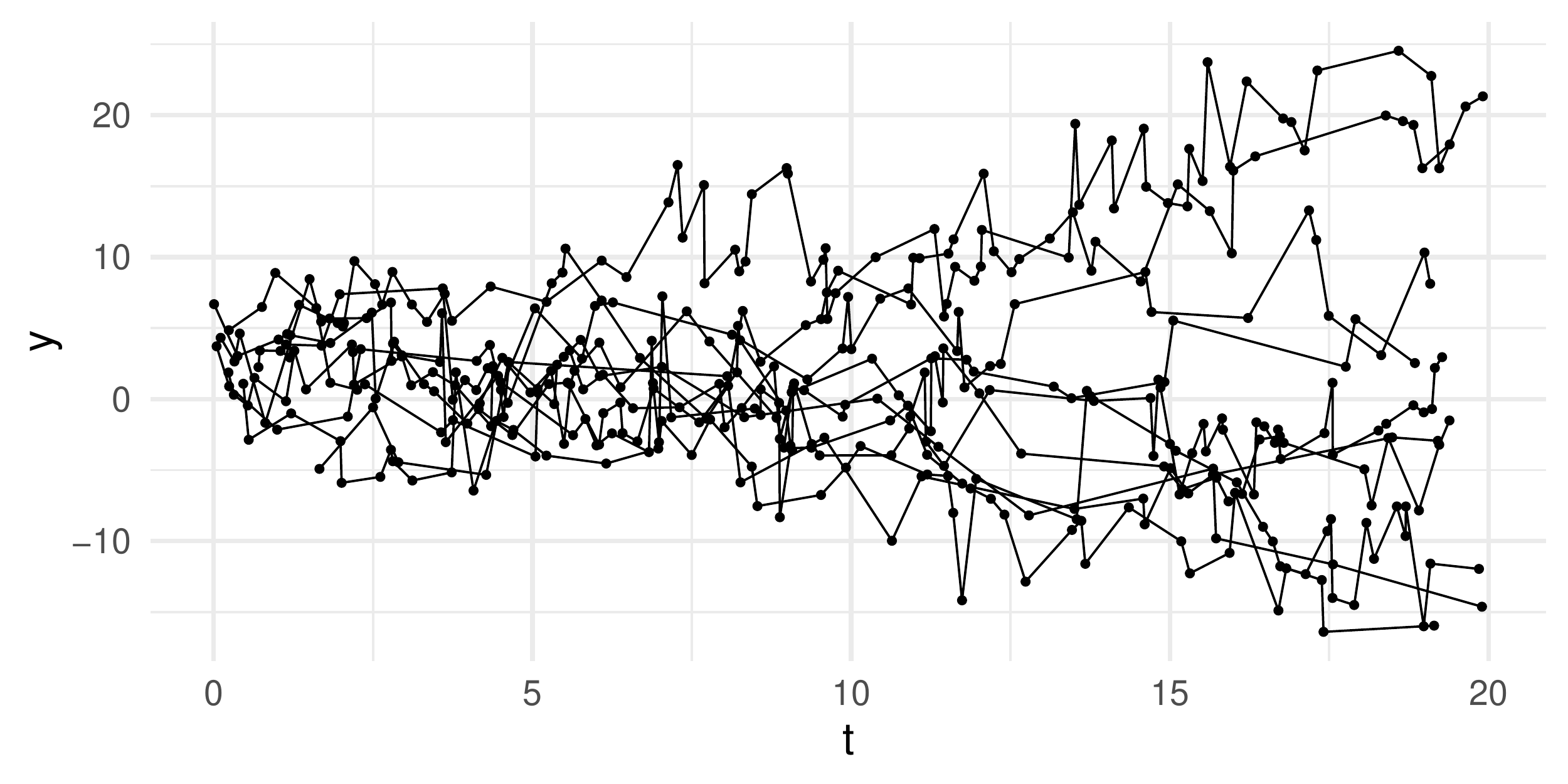} 
	 	\caption{Spaghetti plot of the unbalanced longitudinal data from the second study, 10 individuals are considered with a total of 441 observations. } \label{fig:spaghetti_plots_sim2}
\end{figure}

We consider a scenario defined by an unbalanced set of simulated longitudinal data generated (see Figure \ref{fig:spaghetti_plots_sim2})  by  a LMM   which includes for each individual ($i$, $i=1,\ldots,10$) a common intercept $ \beta_0 $, an individual random  slope   $b_{1i}$,   with     $(b_{1i} \mid \sigma_1) \sim \mathcal{N}(0, \sigma_1^2)$, and   an autoregressive latent element $\boldsymbol w_i=(w_i(t_{i1}), \ldots, w_i(t_{in_{i}}))^{\prime}$ in the conditional mean   as   defined in (\ref{eqn:normal2}) with parameters $\sigma^2_{w}$ and $\rho$, and a common conditional variance $\sigma^2$. This model, which we will call $\mathcal M_1$, can be written as: 

\begin{align} 
 \mathcal M_1: &  \,\,f(\boldsymbol y_{i} |\btheta_1, \bphi_1) = \mathcal{N}(\boldsymbol{1} \beta_0  + \boldsymbol t_i b_{1i} + \boldsymbol w_i, \,\boldsymbol{1}\sigma^2 ).  \label{eqn:model_simulation2a}
\end{align}
 
 We simulate from this model using the values $\beta_0=2$,  $\sigma_1 = 0.5$,  $\sigma = 2$, $\rho=0.8$, and $\sigma_w=1.5$  for the parameters and hyperparameter of the model. Observation times  of the response variable for each individual $i$,  $\boldsymbol t_i=(t_{i1}, \ldots, t_{in_i})^{\prime}$, were generated in a doubly random manner, i.e., both the number of observations (between 10 and 70) and all times when each observation is recorded (between 0 and 20).  The total number of observations registered for the 10 individuals is 441.

We compare $\mathcal{M}_1$ with two alternative models  $\mathcal M_2$ and $\mathcal M_3$ defined as follows:
\begin{align}
 \mathcal M_2: & \,\, f(\boldsymbol y_{i} |\btheta_2, \bphi_2) = \mathcal{N}(\boldsymbol{1} \beta_0  + \boldsymbol 1 b_{0i}+ \boldsymbol t_i b_{1i}, \,\boldsymbol{1}\sigma^2),  \nonumber\\
 \mathcal M_3 : & \,\, f(\boldsymbol y_{i} |\btheta_3, \bphi_3) = \mathcal{N}(\boldsymbol{1} \beta_0   + \boldsymbol t_i b_{1i}, \, \Sigma_{AR}) \label{eqn:model_simulation2b}.  
\end{align} 

Both competing models $\mathcal M_2$ and $\mathcal M_3$ are somewhat simpler than $\mathcal M_1$. Model $\mathcal M_2$ exchanges the autoregressive term for an individual random effect in the intercept $b_{0i}$,   with     $(b_{0i} \mid \sigma_0) \sim \mathcal{N}(0, \sigma_0^2)$, and model $\mathcal M_3$ with only a random slope effect but with an autoregreesive element in the measurement errors.

We complete the Bayesian models by eliciting a prior distribution for the subsequent parameters and hyperparameters. We assume
prior independence among them and select a uniform distribution $\mbox{U}(0, 10)$ for the standard deviation parameters $\sigma$, $\sigma_0$, $\sigma_1$ and $\sigma_w$, and a $\mbox{U}(-1, 1)$ for the autoregressive parameter $\rho$. The normal distribution $\mbox{N}(0, 10^2)$ is  chosen for the common intercept $\beta_0$.

\begin{table}[H]
 \centering
\caption{Mean (standard deviation) of the ten replicates of the approximate log evidence for  models  in (\ref{eqn:model_simulation2a}) and (\ref{eqn:model_simulation2b}) computed by means of the power posterior method.}\label{tab:res_sim2}
\begin{tabular}{ccc}
\noalign{\hrule height 1pt}
               $\mathcal M_1$                               & $\mathcal M_2$                              & $\mathcal M_3$                               \\ \cline{1-3}
 \multicolumn{1}{r}{ $-1123.54 \,(0.48)$} & \multicolumn{1}{r}{$-1158.45\,(1.17)$} & \multicolumn{1}{r}{$-1131.04 \,(1.13)$} \\
\noalign{\hrule height 1pt}
\end{tabular}
\end{table}

 Table \ref{tab:res_sim2} shows, for each model, the mean and the standard deviation of the 10 replicates of  the approximate   logarithmic evidence.  Model $\mathcal M_1$, which is the    true sampling model, presents the highest evidence ($-1123.54$), followed by $\mathcal M_3$ ($-1131.04$). The lowest marginal likelihood value ($-1158.45$) is for model $\mathcal M_2$. Note that the  variability associated with the computation of the model predictive probability is higher than that of the previous simulation study  due to the complexity of these models.

\section{Sardine landings in the Mediterranean Sea}

 Small pelagic fish species are key elements of the Mediterranean pelagic ecosystem \citep{albo2015feeding}.   Fluctuations in populations of these species can provide serious ecological and socio-economic consequences \citep{pennino2020current}.   

Catches in the Mediterranean Sea  are dominated by small pelagics  representing nearly 49\% of the harvest \citep{ramirez2021sos}. Among them, the European sardine ({\sl Sardina pilchardus}, Walbaum, 1792) is one of the most commercial species   which has also shown the highest over-exploitation rates in the 20 years \citep{coll2008food}. Mediterranean fisheries are highly diverse and geographically varied, not only because of the existence of different marine environments, but also because of different socio-economic situations, and fisheries status \citep{pennino2017analysis}.

\begin{figure}[H]
	\centering
	  \hspace*{0cm}\includegraphics[width=130mm]{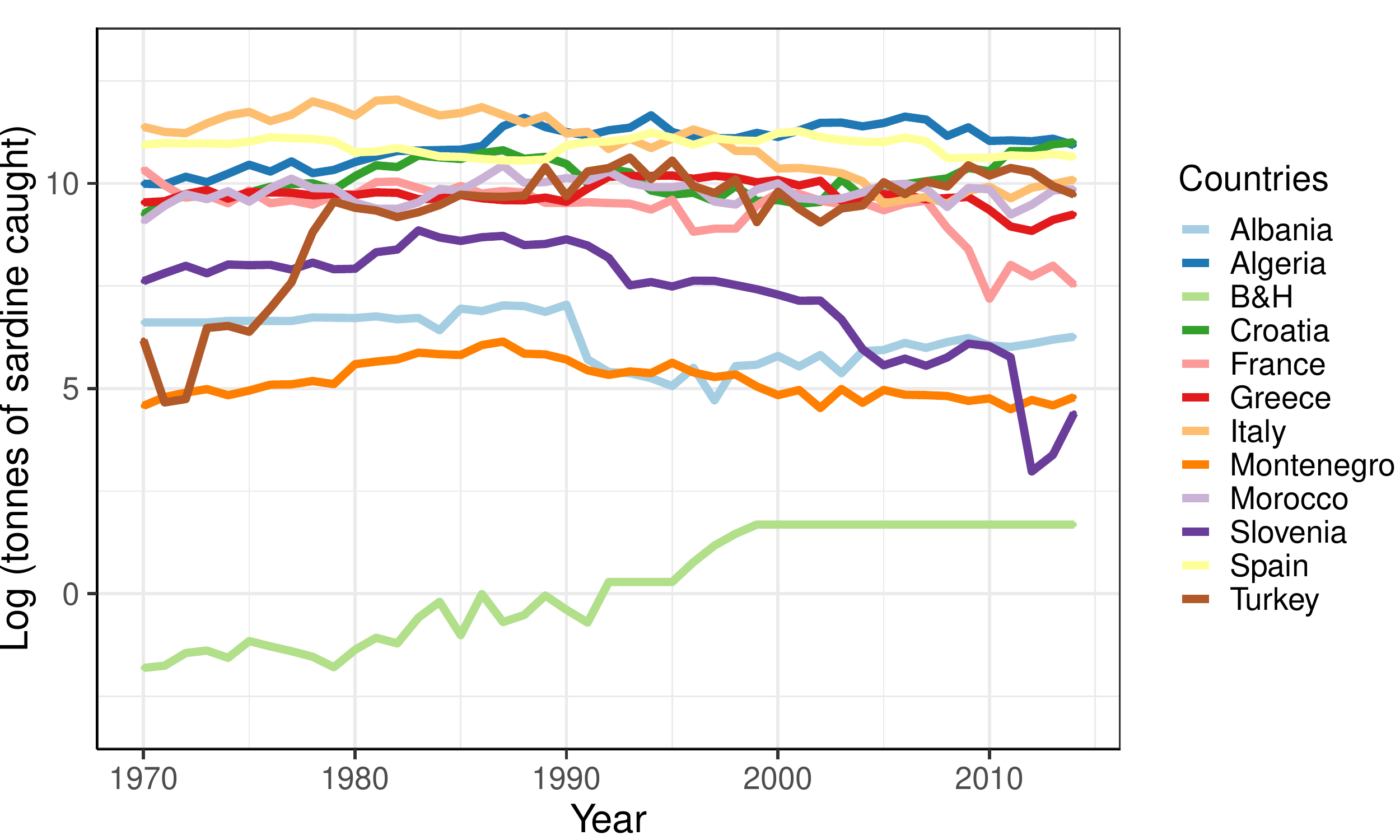} 
	 	\caption{Annual log tons of European sardine ({\sl Sardina pilchardus}) caught per country, from 1970 to 2014.} \label{fig:spaghetti_plots}
\end{figure}

Landing data (tonnes) of the European sardine caught by Mediterranean countries were extracted from  \textsl{Sea Around Us} \citep{zeller2016} from 1970 to 2014,  directly to the online databases (www.seaaroundus.org).
Countries participating in the study were Albania, Algeria, Bosnia and Herzegovina (B\&H), Croatia, France, Greece, Italy, Montenegro, Morocco, Slovenia, Spain and Turkey. 
Data from countries recognised as sovereign after 1970 by the international community (B\&H, Croatia, Montenegro, and Slovenia) were imputed based on information from Exclusive Economic Zones \citep{zeller2016}. This online database is derived mainly from FAO global fisheries catch statistics, complemented by the statistics of various international and national agencies, and reconstructed datasets. It is important to note that the data we are working with come from official fisheries and probably many more sardines are actually caught than those our data reflect (i.e., due illegal, unreported and unregulated fishing). Finally,  a logarithmic transformation was applied to the landing data set in order to approach the normality assumption. Figure \ref{fig:spaghetti_plots} shows the temporal pattern of the logarithm of the tonnes of sardines caught per country. 

The amount of fish caught at the beginning of the study is very variable among the different countries. B\&H   is initially well below the level of the other countries. The temporal evolution of this quantity  in most of the countries seems to be rather stable, although a slightly increasing trend can be seen in most countries, especially in B\&H and Turkey. Slovenia's behaviour in recent years has been different from that of the others, with a decreasing trend in the number of catches in the recent years. The annual changes that we observe in the same country with respect to the previous years are generally not very large, so it would seem reasonable to consider an autoregressive component when modelling the evolution of the sardine landings.

\subsection{Modelling of sardine fisheries in the Mediterranean Sea}

Let $y_{it}$ be the logarithm of the total tonnage of sardines  caught in   country $i$ ($i=1,\ldots,12$) during year $t$ ($t=0,\dots,44$). Calendar time is the natural time scale of the study and $t=0$ corresponds to 1970, the first year of the study. 

We consider  three  longitudinal models, $\mathcal M_1, \mathcal M_2, \text{ and } \mathcal M_3$,    to assess  the dynamics of the official sardine fishery carried out by country $i$ in the Mediterranean Sea from 1970 to 2014. They were defined as follows
\begin{align} 
 \mathcal M_1: &  \,\,f(\boldsymbol y_{i} |\btheta_1, \bphi_1) = \mathcal{N}(\boldsymbol{1} \,\beta_0  + \boldsymbol 1 \,b_{0i}+ \boldsymbol t \, b_{1i}, \,\boldsymbol{1}\sigma^2) , \nonumber \\
 \mathcal M_2: & \,\, f(\boldsymbol y_{i} |\btheta_2, \bphi_2) = \mathcal{N}(\boldsymbol{1} \, \beta_0  + \boldsymbol t \, b_{1i}, \,\bSigma_{AR}) ,  \nonumber \\
 \mathcal M_3: & \,\, f(\boldsymbol y_{i} |\btheta_3, \bphi_3) = \mathcal{N}(\boldsymbol{1}\, \beta_0 + \boldsymbol t \, b_{1i} + \boldsymbol{w}_i, \,\boldsymbol{1}\sigma^2) , \label{eqn:models_sardines}
\end{align}
 
\noindent where in all models $\boldsymbol{1}$ is now a vector of ones of dimension 45, $\boldsymbol t=(0, 1, \ldots, 44)^{\prime}$,  $\beta_0$ is a common intercept and $\sigma^2$ a common variance.   Individual random effects $b_{0i}$ and random slopes $b_{1i}$  are mutually independent and conditionally  normally distributed  as $(b_{0i} \mid \sigma_0^2) \sim \mathcal{N}(0, \sigma_0^2)$ and $(b_{1i} \mid \sigma_1^2) \sim \mathcal{N}(0, \sigma_1^2)$. The elements of the variance-covariance matrix in model $\mathcal M_2$ are as in (\ref{eqn:covariance}), and the autoregressive element  $\boldsymbol{w}_i$ of model
$\mathcal M_2$  is defined as in (\ref{eqn:normal2}).

 We complete the Bayesian models in (\ref{eqn:models_sardines}) by eliciting a prior distribution for the subsequent parameters and hyperparameters of each model. In all of them we assume prior independence and  a vague prior scenario. We consider uniform distributions for the standard deviation parameters $\pi(\sigma)=\pi(\sigma_0)=\pi(\sigma_1)=\pi(\sigma_w)=\text{U}(0,5)$ and  $\pi(\rho)=\text{U}(-1,1)$ for the autoregressive parameter in  models $\mathcal M_2$ and $\mathcal M_3$. A normal distribution is considered for the common intercept parameter $\pi(\beta_0)=\mathcal{N}(0, 5^2)$.

The estimation of the log-evidence for all three models in (\ref{eqn:models_sardines})  was derived through the power posterior according to the same strategy followed in the previous section by means of the Gibbs sampling (see Appendix \ref{sec_appendix}).  We considered   two hundred  temperatures for models   $\mathcal M_1$ and $\mathcal M_2$, $80,000$ iterations, and a burn-in of $20,000$ for each temperature value  $\tau_{r}$. However, due to the complexity of the model $\mathcal M_3$, we increased the number of temperature values to $500$, and   obtain the discretisation $0=\tau_0 < \tau_1 < \dots < \tau_{499}=1$, where $\tau_r = (r / 499)^5$ and $r=0,\ldots,499$, as well as to increase the number of iterations to $200,000$ and the burn-in to $50,000$. Table \ref{tab:res_sardines} shows the   mean and the standard deviation of  ten replicates of the  approximate log-evidence of the three models. 

\begin{table}[H]
 \centering
\caption{Mean (standard deviation) of the approximate log evidence for each model  in (\ref{eqn:models_sardines})  by means of the power posterior method.}\label{tab:res_sardines}
\begin{tabular}{ccc}
\noalign{\hrule height 1pt}
               $\mathcal M_1$                               & $\mathcal M_2$                              & $\mathcal M_3$                               \\ \cline{1-3}
 \multicolumn{1}{r}{$-515.57(0.29)$} & \multicolumn{1}{r}{$-191.38(0.14)$} & \multicolumn{1}{r}{$-193.36(0.79)$} \\
\noalign{\hrule height 1pt}
\end{tabular}
\end{table}

The evidence value of the model $\mathcal M_1$ is clearly lower than the other two models $\mathcal M_2$  and $\mathcal M_3$, this indicates that the autoregressive term is relevant in modelling sardine fishing in the Mediterranean Sea. The last two have  similar log-evidence values but if one of them should be selected, this would be $\mathcal M_2$, albeit by a very small margin. Note that the approximate  Bayes factor of model  $\mathcal M_2$ to $\mathcal M_3$ is $7.25$, which gives   evidence in favour of $\mathcal M_2$. but not very strong. An alternative possibility that could be interesting, but beyond our scope in this study, would be to deal with models $\mathcal M_2$ and $\mathcal M_3$ through model averaging procedures \citep{hoeting1999}. 

 \subsection{Longitudinal modeling of European sardine landings}
 
 We will then  carry out a Bayesian statistical analysis of the sardine landings by means of model $\mathcal M_2$ and discuss some of the relevant outputs derived from the subsequent posterior distribution. 
 
 

The   posterior distribution  for  the selected model $\mathcal M_2$ was evaluated through Gibbs sampling, running a chain of 250,000 iterations after a burn-in of 50,000, and thinning the chain at every 250th iteration to reduce its autocorrelation. A summary of the posterior outputs is shown in  Table \ref{tab:post}.

\begin{table}[H]
		\centering
		\caption{Summary of the approximate  posterior distribution of the parameters and hyperparameters  in model $\mathcal M_2$.}\label{tab:post}
		\vspace{0.25cm}
		\begin{tabular}{ccccc}
			\noalign{\hrule height 1pt}
			\multicolumn{1}{l}{} & \multicolumn{1}{c}{mean}    & \multicolumn{1}{c}{sd} & $q_{0.025}$ & $q_{0.975}$ \\ \noalign{\hrule height 1pt}
			\multicolumn{1}{c}{$\beta_{0}$}  & \multicolumn{1}{r}{$7.92$} & \multicolumn{1}{r}{$0.61$} & \multicolumn{1}{r}{$6.48$} & \multicolumn{1}{r}{$9.05$} \\
			\multicolumn{1}{c}{$\sigma_0$}     & \multicolumn{1}{r}{$3.02$} & \multicolumn{1}{r}{$0.95$} & \multicolumn{1}{r}{$0.68$}& \multicolumn{1}{r}{$4.67$}\\
			\multicolumn{1}{c}{$\sigma_1$}     & \multicolumn{1}{r}{$0.03$} & \multicolumn{1}{r}{$0.02$} & \multicolumn{1}{r}{$0.00$}& \multicolumn{1}{r}{$0.06$}\\
				\multicolumn{1}{c}{$\sigma$}     & \multicolumn{1}{r}{$0.31$} & \multicolumn{1}{r}{$0.01$} & \multicolumn{1}{r}{$0.29$}& \multicolumn{1}{r}{$0.34$}\\\multicolumn{1}{c}{$\rho$}     & \multicolumn{1}{r}{$0.97$} & \multicolumn{1}{r}{$0.03$} & \multicolumn{1}{r}{$0.91$}& \multicolumn{1}{r}{$1$}\\ \noalign{\hrule height 1pt}
		\end{tabular}
	\end{table}
	
At the beginning of the study, $t=0$, we notice a substantial common intercept as well as a large heterogeneity among the different countries. The country-specific variability related to the slope is not very large although its magnitude may be relevant due to the magnitude of the $\boldsymbol t$ values. The variability associated with the model is around 0.31 and it is worth noting the high precision of this estimate. Finally, we observe very high and positive values of the correlation coefficient $\rho$  (around 0.97), which would indicate a fairly stable temporal dynamics of the fisheries.

	\begin{figure}[H]
	\centering
	  \hspace*{0cm}\includegraphics[width=120mm]{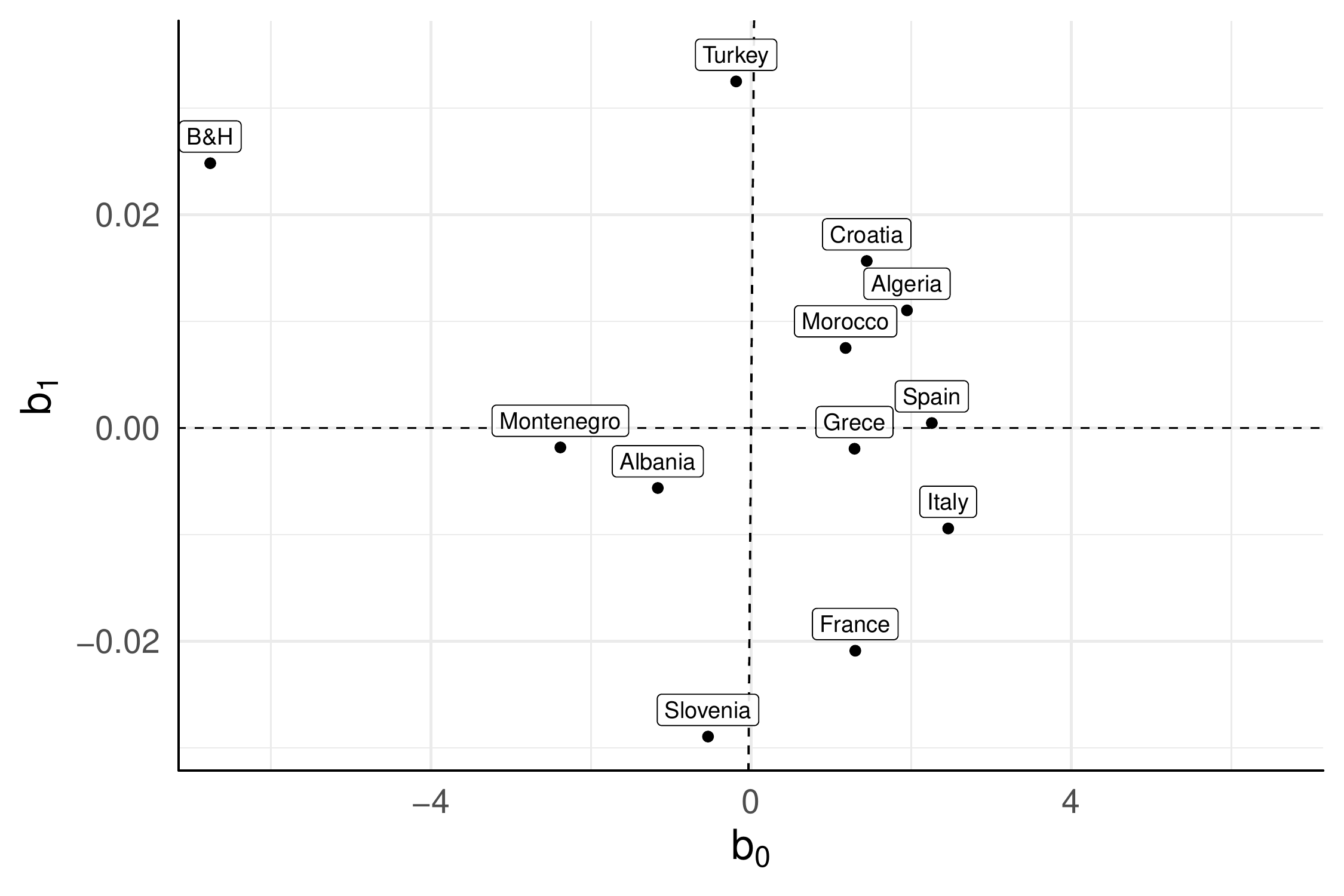} 
	 	\caption{Random intercepts vs. random slopes. Approximate posterior mean for the random intercepts and the slope effects by country according to $\mathcal M_2$.} \label{fig:b0_b1}
\end{figure}

The posterior distribution of the random effects associated with each country provides us with useful information  on the individual country patterns. Figure \ref{fig:b0_b1} shows the posterior mean of the random effect associated with the intercept and the slope of the countries in the study. It can be seen that a group of countries, including Algeria, Croatia and Morocco, had high landing values at the beginning of the time series than the overall average and maintained this superiority over time.  However, Greece and Spain and, to a greater extent, Italy and France also started with above-mean levels of fishing but their growth seems to have slowed down over time. B$\&$H, Turkey, and Slovenia  are countries with quite different dynamics from the rest. B$\&$H started with a level of fishing well below the common mean, but increased the landings during the time-series (possibly as a consequence of the end of the Balkan wars) and now it is above the rest of the countries. The same trend can be observed in Turkey, although in the beginning it was in the mean of the rest of the countries. In contrast, Slovenia's growth has slowed down over the years from  an initial state in the mean of the rest of countries.  
 
\begin{figure}[H]
	\centering
	  \hspace*{0cm}\includegraphics[width=130mm]{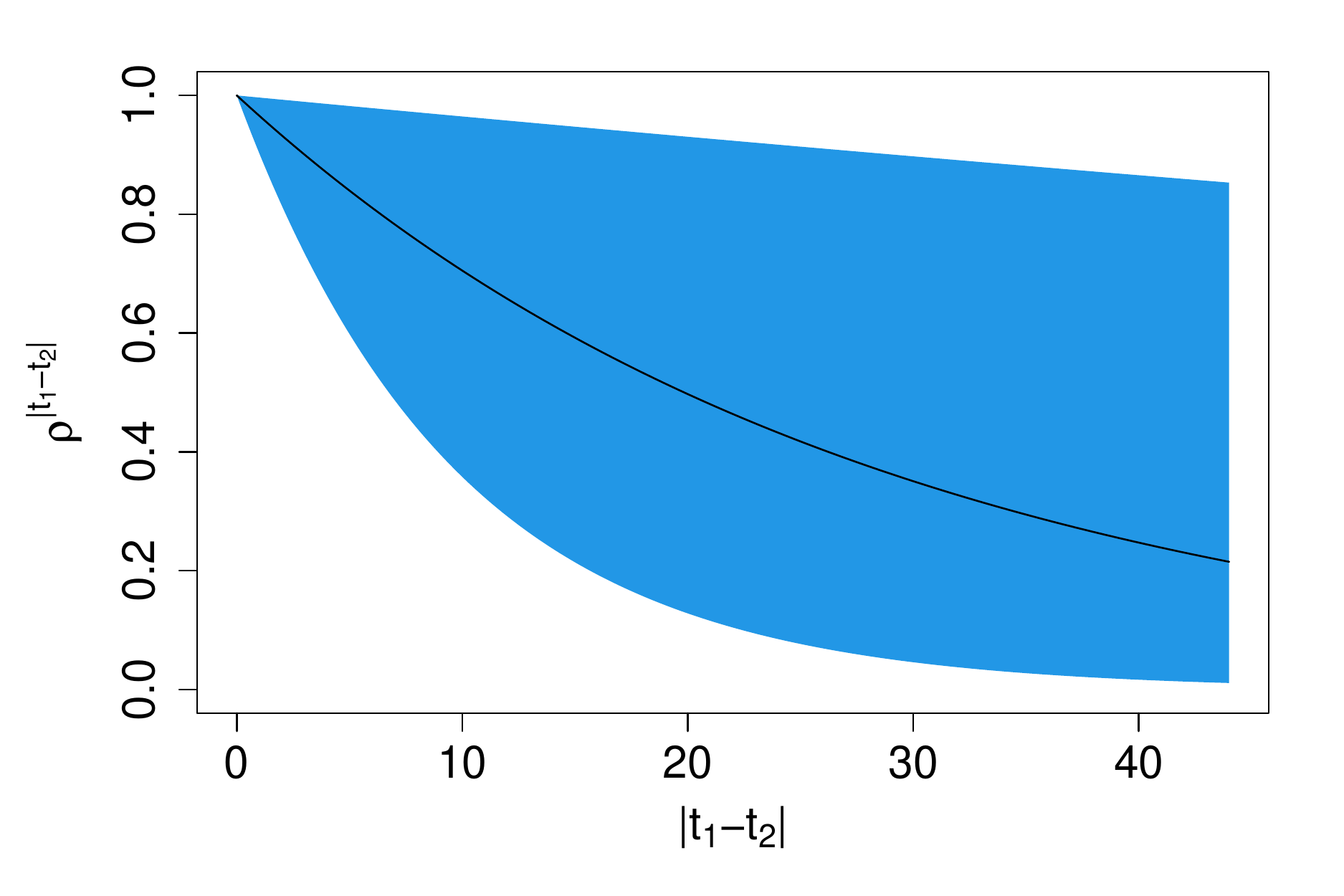} 
	 	\caption{Posterior  mean (line)  and 95\% credible interval (shaded area) of the autocorrelation function  according to the posterior distribution of the autoregressive parameter $\rho$ in $\mathcal{M}_2$}. \label{fig:autocorrelation}
\end{figure}

Finally, we can use the MCMC sample to investigate the autocorrelation function  which depends on the autoregressive parameter $\rho$. Figure \ref{fig:autocorrelation} shows its approximate posterior mean and 95$\%$ credible interval.As expected, the correlation decreases with the increasing of the time lag. However, the uncertainty involved in  the posterior distribution is higher as the lag between the two response variables increases.

The analysis we have presented on sardine landing in the Mediterranean is a small illustration of the potential of Bayesian longitudinal models to analyse the general and individual behaviour of the different elements, in this case countries, of the study. A more detailed study of the problem can be found in \cite{Calvo2020}.

\section{Conclusions}

 The power posterior distribution,  the key concept in the power posterior methodology, allows the computation of  marginal likelihoods by extending the Gibbs sampling process quite  naturally , i. e., by doing Gibbs sampling in each of the power posterior distributions, which are as simple to derive as the posterior distribution. This makes the implementation of this method ideal for Bayesian longitudinal models with different types of random effects and different levels of complexity.  Some variations in the power posterior algorithm can be implemented using importance sampling to avoid sampling from the posterior distribution \citep{xie2011improving} or from distributions close to the prior \citep{fan2011choosing}.  These improvements may slightly reduce the computational cost of the method. In addition, small changes in trapezoidal rule for estimating the evidence on the  on the logarithmic scale  can be applied in order to reduce the bias of the approximation \citep{Friel2012}. 
 
In the two studies based on simulated data that we have examined in the paper, the correct model (i. e., that generated the data) is always selected among competing models with different sources of random variation, demonstrating the efficiency of the power posterior method. Moreover, following this methodology, the conclusion in the analysis of the European Sardine landings in the Mediterranean Sea is that the autoregressive term is relevant in its modelling. Actually, the model that includes two sources of random variation (random effects and autoregressive errors) is that with the highest marginal likelihood value. 
\section*{Acknowledgements}
Gabriel Calvo's research was partially funded by the ONCE Foundation, the Universia Foundation, and the Spanish Ministry of Education and Professional Training, grant FPU18/03101. Carmen Armero and Gabriel Calvo's research was partially funded by the Spanish Research project Bayes$\_$COCO (PID2019-106341GB-I00) from the Ministry of Science and Innovation Grant. MGP's research was funded by the project IMPRESS (RTI2018-099868-B-I00), ERDF, Ministry of Science, Innovation and Universities - State Research Agency. Luigi Spezia's research was funded by the Scottish Government's Rural and Environment Science and Analytical Services Division. Comments from Glenn Marion improved the quality of the final paper.

\bibliographystyle{apalike}
\bibliography{Gabriel}

\section{Appendix. Complete posterior conditional distribution associated of the power posteriors}\label{sec_appendix}
In this appendix the posterior conditional densities needed  to compute the power posteriors via the Gibbs sampling are listed. The conditional densities of the three most general models ($\mathcal M_1$, $\mathcal M_2$, $\mathcal M_3$ of the second simulation study) are fully described. Variability associated with the normal distributions are expressed in terms of the variance.

\subsection{Study 2: an unbalanced longitudinal data with serial correlation}
\subsubsection{Model $\mathcal M_1$}

\begin{itemize}
\item $\pi(\beta_0 | \boldsymbol{y}_{obs}, \bb, \boldsymbol{w}, \sigma, \tau) = \mathcal{N}(\frac{10^2 \tau \sum_{i = 1}^N \sum_{j=1}^{n_i} (y_{ij}- (b_{1i}t_{ij} + w_{ij}))}{\sigma^2 + 10^2 \tau \sum_{i=1}^N n_i}, \frac{10^2\sigma^2}{\sigma^2 + 10^2 \tau \sum_{i=1}^N n_i})$,

\item $\pi(b_{1i} | \boldsymbol{y}_{obs}, \beta_0, \boldsymbol{w}_i, \sigma, \sigma_1, \tau) = \mathcal{N}(\frac{\sigma_1^2 \tau \sum_{j=1}^{n_i} t_{ij}(y_{ij}- (\beta_0 + w_{ij}))}{\sigma^2 + \sigma_1^2 \tau \sum_{j=1}^{n_i} t_{ij}^2}, \frac{\sigma_1^2\sigma^2}{\sigma^2 + \sigma_1^2 \tau \sum_{j=1}^{n_i} t_{ij}^2})$,

\item $\pi(w_{i1} | \boldsymbol{y}_{obs}, \beta_0, \boldsymbol{w}_{- i1}, \rho, \sigma, \sigma_w, \tau) = \mathcal{N}(\frac{\sigma_w^2 \tau (y_{i1}- (\beta_0 + b_{1i}t_{i1})) + \sigma^2\rho w_{i2}}{\sigma^2 + \sigma_w^2 \tau}, \frac{\sigma_w^2\sigma^2}{\sigma^2 + \sigma_w^2 \tau})$,

\item $\begin{aligned}[t]\pi(w_{ij} &| \boldsymbol{y}, \beta_0, \boldsymbol{w}_{- ij}, \rho, \sigma, \sigma_w, \tau) = \\
& \mathcal{N}(\frac{\sigma_w^2 \tau (y_{ij}- (\beta_0 + b_{1i}t_{ij})) + \sigma^2\rho (w_{ij-1} + w_{ij+1})}{\sigma^2 (1+\rho^2)+ \sigma_w^2 \tau}, \frac{\sigma_w^2\sigma^2}{\sigma^2(1+\rho^2) + \sigma_w^2 \tau}), \end{aligned}$ 

\item $\pi(w_{in_i} | \boldsymbol{y}_{obs}, \beta_0, \boldsymbol{w}_{- in_i}, \rho, \sigma, \sigma_w, \tau) = \mathcal{N}(\frac{\sigma_w^2 \tau (y_{in_i}- (\beta_0 + b_{1i}t_{in_i})) + \sigma^2\rho w_{in_i-1}}{\sigma^2 + \sigma_w^2 \tau}, \frac{\sigma_w^2\sigma^2}{\sigma^2 + \sigma_w^2 \tau})$,

\item $\pi(\sigma | \boldsymbol{y}_{obs}, \beta_0, \bb, \boldsymbol{w}, \tau) \propto \frac{1}{\sigma^{\tau \sum_{i=1}^N n_i}} \text{exp} \big(\frac{- \tau \sum_{i = 1}^N \sum_{j=1}^{n_i} (y_{ij}- (\beta_0 + b_{1i}t_{ij} + w_{ij}))^2}{2\sigma^2} \big)\text{U}(0,10)$,

\item $\pi(\sigma_1 | \boldsymbol{y}_{obs}, \bb) \propto \frac{1}{\sigma_1^{N}} \text{exp} \big(\frac{- \sum_{i = 1}^N b_{1i}^2}{2\sigma_1^2} \big)\text{U}(0,10)$,

\item $\begin{aligned}[t]\pi(\sigma_w | \boldsymbol{y}_{obs}, \rho, \boldsymbol{w}) \propto 
\frac{(1-\rho^2)^{N/2}}{\sigma_w^{\sum_{i=1}^N n_i}}
& \text{exp} \big(\frac{- (1 - \rho^2)\sum_{i = 1}^N  w_{i1}^2}{2\sigma_w^2} \big)\times\\
&\text{exp} \big(\frac{- \sum_{i = 1}^N \sum_{j=1}^{n_i} (w_{ij} - \rho w_{ij-1})^2}{2\sigma_w^2} \big)
\text{U}(0,10),\end{aligned}$

\item $\begin{aligned}[t]\pi(\rho | \boldsymbol{y}_{obs}, \sigma_w, \boldsymbol{w}) \propto 
\frac{(1-\rho^2)^{N/2}}{\sigma_w^{N}}
& \text{exp} \big(\frac{- (1 - \rho^2)\sum_{i = 1}^N  w_{i1}^2}{2\sigma_w^2} \big)\times\\
&\text{exp} \big(\frac{- \sum_{i = 1}^N \sum_{j=1}^{n_i} (w_{ij} -\rho w_{ij-1})^2}{2\sigma_w^2} \big)
\text{U}(-1,1).\end{aligned}$
\end{itemize}

\subsubsection{Model $\mathcal M_2$}
\begin{itemize}
\item $\pi(\beta_0 | \boldsymbol{y}_{obs}, \bb, \sigma, \tau) = \mathcal{N}(\frac{10^2 \tau \sum_{i = 1}^N \sum_{j=1}^{n_i} (y_{ij}- (b_{0i} + b_{1i}t_{ij}))}{\sigma^2 + 10^2 \tau \sum_{i=1}^N n_i}, \frac{10^2\sigma^2}{\sigma^2 + 10^2 \tau \sum_{i=1}^N n_i})$,

\item $\pi(b_{0i} | \boldsymbol{y}_{obs}, \beta_0, \bb_{1}, \sigma, \sigma_0, \tau) = \mathcal{N}(\frac{\sigma_0^2 \tau \sum_{j=1}^{n_i} (y_{ij}- (\beta_0 + b_{1i}t_{ij}))}{\sigma^2 + \sigma_0^2 \tau n_i}, \frac{\sigma_0^2\sigma^2}{\sigma^2 + \sigma_0^2 \tau n_i})$,

\item $\pi(b_{1i} | \boldsymbol{y}_{obs}, \beta_0, \bb_0, \sigma, \sigma_1, \tau) = \mathcal{N}(\frac{\sigma_1^2 \tau \sum_{j=1}^{n_i} t_{ij}(y_{ij}- (\beta_0 + b_{0i}))}{\sigma^2 + \sigma_1^2 \tau \sum_{j=1}^{n_i} t_{ij}^2}, \frac{\sigma_1^2\sigma^2}{\sigma^2 + \sigma_1^2 \tau \sum_{j=1}^{n_i} t_{ij}^2})$,

\item $\pi(\sigma | \boldsymbol{y}_{obs}, \beta_0, \bb, \boldsymbol{w}, \tau) \propto \frac{1}{\sigma^{\tau \sum_{i=1}^N n_i}} \text{exp} \big(\frac{- \tau \sum_{i = 1}^N \sum_{j=1}^{n_i} (y_{ij}- (\beta_0 + b_{0i} + b_{1i}t_{ij}))^2}{2\sigma^2} \big)\text{U}(0,10)$,

\item $\pi(\sigma_0 | \boldsymbol{y}_{obs}, \bb_0) \propto \frac{1}{\sigma_0^{N}} \text{exp} \big(\frac{- \sum_{i = 1}^N b_{0i}^2}{2\sigma_0^2} \big)\text{U}(0,10)$,

\item $\pi(\sigma_1 | \boldsymbol{y}_{obs}, \bb_1) \propto \frac{1}{\sigma_1^{N}} \text{exp} \big(\frac{- \sum_{i = 1}^N b_{1i}^2}{2\sigma_1^2} \big)\text{U}(0,10)$.
\end{itemize}

\subsubsection{Model $\mathcal M_3$}
\begin{itemize}
\item $\begin{aligned}[t]\pi(& \beta_0 |\boldsymbol{y}_{obs}, \rho, \bb_1,  \sigma, \tau) = \\
 & \Scale[1]{ \mathcal{N}(\frac{10^2 \tau \big[ (1-\rho^2)\sum_{i = 1}^N (y_{i1} - b_{1i}t_{i1}) - (1-\rho)  \sum_{i = 1}^N \sum_{j=2}^{n_i} (\rho y_{ij-1} - y_{ij} + (t_{ij}- \rho t_{ij-1})b_{1i}) \big]}{\sigma^2 + 10^2 \tau [ (1-\rho^2)N + (1-\rho)^2 \sum_{i=1}^N (n_i-1) ]}},\\
& \frac{10^2\sigma^2}{\sigma^2 + 10^2 \tau [ (1-\rho^2)N + (1-\rho)^2 \sum_{i=1}^N (n_i-1) ]}),\end{aligned}$

\item $\begin{aligned}[t]\pi(& b_{1i} | \boldsymbol{y}_{obs}, \beta_0, \sigma, \sigma_1, \tau) = \\
&\mathcal{N}(\frac{\sigma_1^2 \tau \big[ t_{i1}(y_{i1} - \beta_0) - \sum_{j=2}^{n_i} (t_{ij} - \rho t_{ij-1})(\rho y_{ij-1}- y_{ij} + (1-\rho) \beta_0)\big]}{\sigma^2 + \sigma_1^2 \tau [(1-\rho^2)t_{i1}^2 + \sum_{j=2}^{n_i} (t_{ij}-\rho t_{ij-1})^2]},\\
 &\frac{\sigma_1^2\sigma^2}{\sigma^2 + \sigma_1^2 \tau [(1-\rho^2)t_{i1}^2 + \sum_{j=2}^{n_i} (t_{ij}-\rho t_{ij-1})^2]}),\end{aligned}$

\item $\begin{aligned}[t]f(\sigma | &\boldsymbol{y}_{obs}, \beta_0, \rho, \bb_1, \tau) \propto \Scale[1]{\frac{1}{\sigma^{\tau \sum_{i=1}^N n_i}} \text{exp} \big(-\frac{ \tau (1-\rho^2) \sum_{i=1}^N (y_{i1}-(\beta_0+b_{1i}t_{i1}))^2}{2\sigma^2} \big) \times}\\
 &\Scale[1]{ \text{exp} \big(-\frac{ \tau \sum_{i = 1}^N \sum_{j=2}^{n_i} (y_{ij}- [(1-\rho)\beta_0 + b_{1i}(t_{ij}-\rho t_{ij-1}) + \rho y_{ij-1}])^2}{2\sigma^2} \big)\text{U}(0,10)},\end{aligned}$

\item $\pi(\sigma_1 | \boldsymbol{y}_{obs}, \bb_1) \propto \frac{1}{\sigma_1^{N}} \text{exp} \big(\frac{- \sum_{i = 1}^N b_{1i}^2}{2\sigma_1^2} \big)\text{U}(0,10)$,

\item $\begin{aligned}[t]\pi(&\rho | \boldsymbol{y}_{obs}, \beta_0, \bb_1, \tau) \propto \Scale[1.1]{ \frac{(1-\rho^2)^{N/2}}{\sigma^{\tau \sum_{i=1}^N n_i}} \text{exp} \big(-\frac{ \tau (1-\rho^2) \sum_{i=1}^N (y_{i1}-(\beta_0+b_{1i}t_{i1}))^2}{2\sigma^2} \big) \times }\\
  & \Scale[1.1]{ \text{exp} \big(-\frac{ \tau \sum_{i = 1}^N \sum_{j=1}^{n_i} (y_{ij}- [(1-\rho)\beta_0 + b_{1i}(t_{ij}-\rho t_{ij-1}) + \rho y_{ij-1}])^2}{2\sigma^2} \big)\text{U}(-1,1)}.\end{aligned}$
\end{itemize}

\end{document}